\documentclass[prb ,nobibnotes,aps, superscriptaddress, reprint, amsmath,amssymb, showpacs]{revtex4-1}
\usepackage{amsmath,latexsym,psfrag}
\usepackage[pdftex]{graphicx}
\usepackage{epstopdf}
\usepackage{natbib}
\usepackage{array}
\usepackage{dcolumn}
\usepackage{bm}
\usepackage{soul}  

\usepackage[%
colorlinks=true,
urlcolor=blue,
linkcolor=blue,
citecolor=blue
]{hyperref}
\usepackage{gensymb}
\begin{document}

	\title{First-principles study of two-dimensional electron and hole gases at the head-to-head 
		and tail-to-tail 180$^\circ$ 
		domain walls in PbTiO$_{3}$ ferroelectric thin films}
	
	\author{ James Sifuna}
  \email{sifunajames@gmail.com}
	
	\affiliation{ Departamento de Ciencias de la Tierra y
		F\'{\i}sica de la Materia Condensada, Universidad de Cantabria,
		Cantabria Campus Internacional,
		Avenida de los Castros s/n, 39005 Santander, Spain.}
	
	\affiliation{ Department of Natural Science, 
		The Catholic University of Eastern Africa,
		62157 - 00200, Nairobi, Kenya.}
	\affiliation{ Materials Modeling Group, 
		Department of Physics and Space Sciences,
		The Technical University of Kenya,
		52428-00200, Nairobi, Kenya.}  
	
	\author{ Pablo Garc\'{\i}a-Fern\'andez}
	\affiliation{ Departamento de Ciencias de la Tierra y
		F\'{\i}sica de la Materia Condensada, Universidad de Cantabria,
		Cantabria Campus Internacional,
		Avenida de los Castros s/n, 39005 Santander, Spain.}
	
	\author{George S. Manyali}
	\affiliation{
		Computational and Theoretical Physics Group, Department of Physical sciences,
		Kaimosi Friends University College,
		385-50309, Kaimosi, Kenya.}
	
	\author{George Amolo}
	\affiliation{ Materials Modeling Group, 
		Department of Physics and Space Sciences,
		The Technical University of Kenya,
		52428-00200, Nairobi, Kenya.}

	\author{ Javier Junquera }
	\email{javier.junquera@unican.es}
	\affiliation{ Departamento de Ciencias de la Tierra y
		F\'{\i}sica de la Materia Condensada, Universidad de Cantabria,
		Cantabria Campus Internacional,
		Avenida de los Castros s/n, 39005 Santander, Spain.}
	
	\date{\today}
	
	\begin{abstract}
		We study from first-principles the structural and electronic properties of head-to-head (HH)
		and tail-to-tail (TT) 180$^\circ$ domain walls in isolated free-standing PbTiO$_{3}$ slabs.
		For sufficiently thick domains ($n$ = 16 unit cells of PbTiO$_{3}$),
		a transfer of charge from the free surfaces to the 
		domain walls to form localized electron (in the HH) and hole (in the TT) gases
		in order to screen the bound polarization charges is observed.
		The electrostatic driving force behind this electronic reconstruction is 
		clearly visible from the perfect match between the smoothed free charge densities
		and the bound charge distribution, computed from a finite difference of the polarization profile
		obtained after the relaxation of the lattice degrees of freedom.
		The domain wall widths, of around six unit cells, are larger than in the conventional 
		180$^\circ$ neutral configurations.
		Since no oxygen vacancies, defects or dopant atoms are introduced in our simulations,
		all the previous physical quantities are the intrinsic limits of the system.
		Our results support the existence of an extra source of charge at the domains walls
		to explain the enhancement of the conductivity observed in some domains walls of prototypical,
		insulating in bulk, perovskite oxides.
	\end{abstract}
	

	\maketitle
	\section{INTRODUCTION}
	\label{sec:introduction}
	
	Domain walls (DWs) in oxide ferroelectric materials,
	defined as nanometer-scale interfaces that separate regions 
	with different orientations of the spontaneous electric polarization,
	have attracted a lot of attention during the last few years due to the
	novel functional properties they might display, 
	radically departing from those observed in the parent 
	materials~\cite{Catalan-12}.
	One of these surprises comes from the observation of a room-temperature 
	electronic conductivity at ferroelectric DWs in a thin-film 
	insulating multiferroic BiFeO$_{3}$~\cite{Seidel-09}.
	Following this work, Guyonnet and co-workers~\cite{Guyonnet-11} demostrated
	that 180$^\circ$ DWs in much simpler, purely ferroelectric
	tetragonal perovskite Pb(Zr$_{0.2}$Ti$_{0.8}$)O$_{3}$ thin films were also
	conducting, suggesting that this phenomenon is more general than initially envisaged.
	Indeed, the presence of a conductive domain wall is not restricted
	to thin films, but equally applies to millimeter‐thick wide‐bandgap
	single crystals, such as LiNbO$_3$~\cite{Schroder-12} or WO$_{3}$~\cite{Kim-10}.
	The microscopic origin of the conduction is controversial. 
	It was first ascribed to structural changes at the DW that produced
	a polarization discontinuity, leading to steps in the
	electrostatic potential, and a concomitant lowering of 
	the band gaps~\cite{Seidel-09}.
	More recently, it has been suggested that the conductivity 
	could be linked with {\it extrinsic factors}, such as 
	oxygen vacancies~\cite{Farokhipoor-11,Guyonnet-11}.
	
	But other sources for a metallic type-conductivity at DWs have been proposed~\cite{Bendnyakov-2018}.
	Sluka {\it et al.}~\cite{Sluka-13,Bednyakov-15} observed an enhancement of conductivity up to 
	10$^9$ times that of the parent matrix in charged DWs of
	the prototypical ferroelectric BaTiO$_{3}$.
	This was attributed to the presence of a stable degenerate electron gas
	that would be formed to screen polarization divergences in head-to-head (HH) and
	tail-to-tail (TT) DWs~\cite{Gureev-11}.
	In conventional 180$^\circ$ domains in tetragonal ferroelectrics, 
	the walls run parallel to the tetragonal polarization axis~\cite{Poykko-99, Meyer-02}.
	In such a configuration, the polarization charge (defined as the dot product of the polarization
	times the unitary vector normal to the DW) vanishes, yielding to a neutral configuration.
	However, tilting of the DWs would lead to polarization-induced
	bound charges that would result in a large and unfavorable electrostatic energy~\cite{Zhang-20}.
	The extreme case, where the polarization points in the same direction as the normal vector to the  wall, 
	leads to the formation of head-to-head DWs (if the polarization
	is directed towards the DW; positive bound charges), or 
	tail-to-tail DWs (if the polarization is directed outwards from the DW; negative bound charges).
	The properties of charged DWs might be very different from those of the 
	neutral configurations.
	Different mechanisms have already been postulated in the literature to explain the screening of the
	polarization bound charges and stabilization
	of HH and TT 180$^{\circ}$ DWs in ferroelectrics.
	The theoretical background for the formation of charged domain walls in proper ferroelectrics
		based on a phenomenological model was established in Ref.~\onlinecite{Bednyakov-15},
		where an analysis of the different factors controlling the energy and the size of charged DWs was carried out.
		A mixed electron/ion screening scenario, with a combination of free equilibrium carriers coming
		from an electronic reconstruction (transfer of charge between the top of the valence band and the
		bottom of the conduction band due to a Zener-like breakdown) and a redistribution 
		of the ionised impurities during the formation of the
		walls, seems the most favorable scenario.
	Phase field simulations support the existence of a degenerate quasi-two-dimensional electron gas,
	localized due to the potential well formed by the polarization charges~\cite{Sluka-13}.
	The application of a Landau model to address the problem of charged DWs in an isolated ferroelectric,
	including the polarization profile, width and formation energy of the domains was undertaken in Ref.~\onlinecite{Gureev-11}.
	From a first-principles perspective, Wu and Vanderbilt~\cite{Wu-06} proposed a periodically repeated chemical ``delta'' doping in 
	PbTiO$_{3}$ superlattices,
	where the Ti atoms at certain layers are replaced by donor (Nb) or acceptor (Sc) atoms drawn from 
	neighboring columns of the Periodic Table.
	Rahmanizadeh {\it et al.}~\cite{Rahmanizadeh-14} suggested for the same material
	(i) the formation of a conducting layer
	at the domain wall in a superlattice geometry; (ii) the appearance of polarons due to the localization of one 
	electron on a Ti ion (requiring for this the addition of an on-site Coulomb repulsion term) also in a superlattice geometry;
	and (iii) the presence of defects with various amounts of concentrations in thin film geometries.
	But in all the former first-principles simulations, the presence of substitutional atoms 
	at one or the two HH and TT domain walls in the superlattices, or
	a varying defect concentration at the tails in the thin film geometry were assumed, 
	and this fully determines the electric displacement, and therefore the polarization, within the domain
	as was shown in Ref.~\onlinecite{Stengel-11.2}.
	Questions like what is the intrinsic critical thickness for the stabilization of the two-dimensional conducting layers for the screening of the polarization charges, the spatial extension of the hole and electron gases,
	or how large the polarization would be, if the HH and TT domain walls are induced in an ideal slab
	are not accessible from the previous computations.
	Similar issues regarding the screening of polarization discontinuities with electronic reconstructions have been
	the subject of intense studies in the celebrated LaAlO$_{3}$/SrTiO$_{3}$ interfaces 
	(Ref.~\onlinecite{Stengel-11.2} and references therein), and theoretically in KNbO$_{3}$/ATiO$_{3}$ (A=Sr,Ba,Pb) 
	superlattices~\cite{Niranjan-09,Garcia-Fernandez-13}.
	They have been also faced by Aguado-Puente {\it et al.}~\cite{Aguado-Puente-15,Yin-15} 
	in a similar system to ours: an interface between a ferroelectric material like PbTiO$_{3}$ and 
	a dielectric like SrTiO$_{3}$. Above a critical thickness 
	of 16 unit cells, the ferroelectric layers could be stabilized in the out-of-plane monodomain configuration 
	due to the electrostatic screening provided by the free carriers. 
	We wonder whether the same results hold when the
	dielectric is replaced by an opposite domain, doubling the amount of polarization charge at the DW.
	
	The rest of the paper is organized as follows:
	In Sec.~\ref{sec:computational details} we describe the method on which the simulations are based.
	In Sec.~\ref{sec:results} we present the results on the polarization profiles
	(Sec.~\ref{sec:polarization}), the electronic properties including the 
	layer-by-layer projected density of states (Sec.~\ref{sec:pdos}),
	and the transfer of charge that yields to the formation of the two-dimensional hole
	and electron gases (Sec.~\ref{sec:charge carriers}), together with the coupling between
	the electronic properties (density of free charge) and the structural properties
	(divergence of the polarization).
	Future perspectives are summarized in Sec.~\ref{sec:conclusions}.
	
	\begin{figure} [h]
		\centering
		\includegraphics [width=\columnwidth]{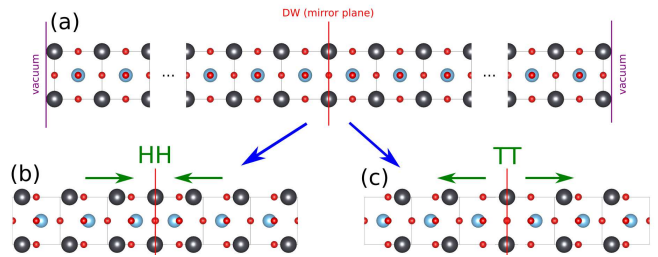}
		\caption {(Color online) schematic representation of the (a) paraelectric phase, (b) head-to-head (HH), and   (c) tail-to-tail (TT) domain walls. Atoms are represented by sph heres: Pb in black, Ti in blue, and O in red.
			Green arrows represent the direction of the polarization.}
		\label{fig:figures}
	\end{figure}

	\section{Methods}
	\label{sec:computational details}
	We carried out simulations of HH and TT
	domain walls in a free standing slab of PbTiO$_{3}$, within a 
	numerical atomic orbital method as implemented in the 
	{\sc Siesta} code~\cite{Soler-02}.
	Exchange  and correlation were treated within the local density approximation 
	(LDA)~\cite{Ceperley-80} to the density functional theory (DFT)~\cite{Hohenberg-64,Kohn-65}. 
	While the use of an on-site Hubbard  term for Ti would allow for a more realistic
		estimation of the band gap, we decided against taking this approach. The reason
		is that such a term would reduce the Ti $3d$-O $2p$
		hybridization, at the origin of the ferroelectric behaviour of PbTiO$_{3}$ and, therefore, 
		also behind the appearance of conductive domains in our simulations. In our tests 
		in bulk PbTiO$_3$ we could not find a U that reproduced the band gap and gave, at the same time, a reasonable value for the material's polarization (indeed, the total polarization falls dramatically for values of U $>$ 2 eV, and
		vanishes for a U $>$ 3 eV).
		Nevertheless, we have to keep in mind that within the bare LDA approach, 
		the band gap is severely understimated, with a concomitant reduction of the critical thickness for the stabilization of the
		conducting layers (in simple models it is proportional to the band gap width).
	
	Core electrons were replaced by {\it ab initio} norm conserving
	pseudopotentials, generated using the
	Troullier-Martins scheme~\cite{Troullier-91} in the
	Kleinman-Bylander fully non-local separable representation~\cite{Kleinman-82}.
	Due to the large overlap between the semicore and valence states,
	the semicore $3s$ and $3p$ electrons of Ti, and the $5d$ electrons of Pb
	were considered as valence electrons
	and explicitly included in the simulations.
	Ti and Pb pseudopotentials were generated scalar relativistically.
	The reference configuration and cutoff radii for each angular momentum shell
	for the pseudopotentials used in this work
	can be found in Ref.~\onlinecite{Junquera-03.2} for Ti and O,
	and in Table~\ref{table:pseudopotentials} for Pb.
	
	\begin{table}
		\caption[ ]{ Reference configuration and cutoff radii (in bohr)
			of the Pb pseudopotential used in our study.
		}
		\begin{center}
			\begin{tabular}{cccc}
				\hline
				\hline
				Reference                         &
				&
				$6s^{2}, 6p^{2}, 5d^{10}, 5f^{0}$  \\
				\hline
				Core radius                       &
				$s$                               &
				2.00                              \\
				&
				$p$                               &
				2.30                              \\
				&
				$d$                               &
				2.00                              \\
				&
				$f$                               &
				1.50                              \\
				Scalar relativistic?              &
				&
				yes                               \\
				\hline
				\hline
			\end{tabular}
		\end{center}
		\label{table:pseudopotentials}
	\end{table}
	
	The one-electron Kohn-Sham eigenstates were expanded in a basis of
	strictly localized numerical atomic orbitals~\cite{Sankey-89,Artacho-99}.
	We used a single-$\zeta$ basis set for the semicore states of
	Ti and Pb, and double-$\zeta$ plus polarization for the valence
	states of all the atoms. For Pb, an extra shell of 5$f$
	orbitals was added.
	All the parameters that define the shape and range of the basis functions
	were obtained by a variational
	optimization of the energy~\cite{Junquera-01} in bulk cubic BaTiO$_{3}$
	(for Ti, and O). For Pb, another optimization was performed in bulk
	PbTiO$_{3}$, frozen in the atomic orbitals of Ti and O to those previously optimized in
	BaTiO$_{3}$~\cite{basis-request}.
	
	The electronic density, Hartree, and exchange-correlation potentials, 
	as well as the corresponding matrix elements
	between orbitals, were calculated in a uniform real-space grid, with an equivalent 
	plane-wave cutoff of 1200 Ry in the representation of charge density.
	For the Brillouin zone integrations we use a Monkhorst-Pack sampling~\cite{Monkhorst-76}
	equivalent to $6 \times 6 \times 6$ in a five atom perovskite unit cell.
	A Fermi-Dirac distribution was chosen for the occupation of the one-particle
	Kohn-Sham electronic eigenstates, with a smearing temperature of 
	250 K for the HH and 500 K for the TT domain structures.
	
	To simulate both domain structures we used a tetragonal $(1\times 1)$ supercell,
	periodically repeated in space, 
	of the type vacuum/(PbO-TiO$_{2}$)$_{2n}$-PbO/vacuum with $n=16$, as shown in 
	Fig.~\ref{fig:figures} 
	(simulations with $n$=12 showed that the system went back to the paraelectric structure
	without stabilizing the metallic gases).
	Both surfaces were terminated with a PbO layer, since 
	this is the only stable surface termination found by first-principles~\cite{Meyer-99}.
	The vacuum thickness was equivalent to roughly eight unit cells of the perovskite.
	A dipole correction was introduced to avoid spurious interaction between periodic images of the slab in the out-of-plane direction.
	In this way, we also enforce that the electric displacement field in vacuum vanishes, and due to the continuity of its normal component at the free-surface, after the 
		electronic reconstruction it exactly amounts within the PbTiO$_{3}$ domain  to
		the surface charge density.
	In order to simulate the effect of the mechanical boundary conditions due to the
	strain imposed by an hypothetical substrate, the in-plane lattice constant was fixed to the
	theoretical equilibrium lattice constant of bulk SrTiO$_{3}$ ($a_{0}= 3.874 $ \AA),
	one of the most common substrates used to grow oxide heterostructures.
	In constrained bulk PbTiO$_{3}$ under this compressive strain, the spontaneous polarization
	amounts to $P_{\rm S} = 77.1$ $\mu$C/cm$^{2}$.
	As the initial point, an ideal structure was defined stacking along the 
	[001] direction $2n$ unit cells of PbTiO$_{3}$ in a tetragonal paraelectric configuration
	(i. e. zero rampling in all the atomic layers) with a tetragonality of $c/a = 1.0325$.
	Then, on top of this paraelectric structure, the PbTiO$_{3}$ atoms were moved in opposite directions
	in the two halves of the slab geometry
	following the displacement pattern of the bulk tetragonal soft mode
	scaled to 62\%. That is the value of the polarization of a meta-stable structure of a monodomain
	PbTiO$_{3}$ slab where the polarization charge can be screened by an electronic
	reconstruction
	and the formation of two-dimensional metallic gases at the
	surfaces~\cite{Aguado-Puente-15}.
	By construction, the central PbO layer is a mirror symmetry plane.
As it was discussed in Ref.~\onlinecite{Meyer-02} 
		there is no indication from first-principles simulations that the energy of the Pb-centered domain wall
		can be lowered by breaking the inversion symmetry.
	Starting from this geometry, a conjugate gradient minimization was performed till the
	maximum component of the force on any atom was smaller than 0.04 eV/\AA.
	
	Once the atomic structure is relaxed and the one-particle density matrix converged,
	in order to compute the density of states, a non-self-consistent calculation was
	carried out with a much denser sampling of $150 \times 150 \times 5$ Monkhorst-Pack mesh.
	
	To establish the notation, we shall call the plane parallel to the interface the $(x,y)$ plane, 
	whereas the perpendicular direction will be referred to as the $z$ axis.

	\section{Results}
	\label{sec:results}
	
	\subsection{Polarization profiles}
	\label{sec:polarization}
	
	After relaxation of the slabs with $n=16$ unit cells in each domain,
	we observe how the two polar configurations are meta-stable, while the most stable phase is the
	centrosymmetric unpolarized structure. 
	A similar result was found in the PbTiO$_{3}$/SrTiO$_{3}$ interfaces ~\cite{Yin-15}.
	The TT configuration is more stable than the HH, as can be seen from the differences in energy 
	with respect the centrosymmetric paraelectric phase shown in Table~\ref{table:fitpol}.
	
\begin{figure} [h]
		\centering
		\includegraphics [width=\columnwidth]{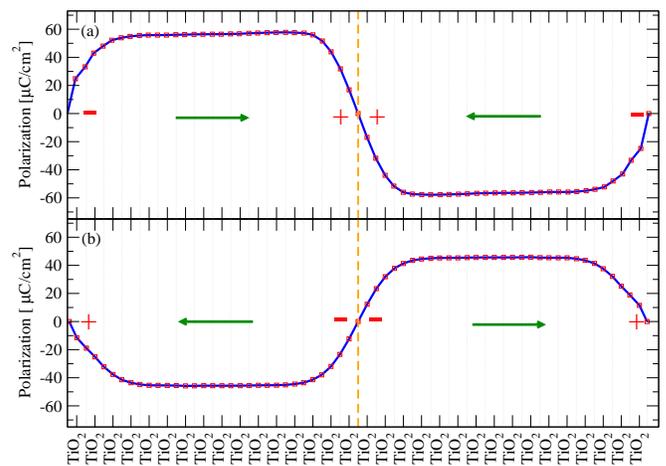}
		\caption {(Color online) Local polarization profiles for (a) head-to-head, and (b) tail-to-tail 180$^\circ$ domain walls
			in PbTiO$_{3}$. Red squares represent the polarization of each layer, computed from ``renormalized'' Born effective charges.
			The solid blue line is a guide to the eye. Vertical dotted grey lines denote the positions of the PbO layers, while the 
			vertical dashed orange line marks the position of the domain wall.
			The flat value of the polarization at the center of each domain amounts to
			$P_{\rm HH}$= 57.26 $\mu$C/cm$^2$ and $P_{\rm TT}$= 45.72 $\mu$C/cm$^2$ 
			for the HH and TT, respectively. Green arrows represent the direction of the
			macroscopic polarization within each domain. The $+$ and $-$ correspond to the sign of the polarization-induced
			charge density at the free surfaces and at the domain wall.}
		\label{fig:layer-by-layer-pol}
	\end{figure}

	In Fig.~\ref{fig:layer-by-layer-pol} we plot the layer-by-layer effective polarization profile in the 
	HH [Fig.~\ref{fig:layer-by-layer-pol}(a)] and TT [Fig.~\ref{fig:layer-by-layer-pol}(b)]
	domain structures for PbTiO$_{3}$. 
	The layer polarizations have been calculated using the approximate expression based
		on the effective Born charge method, where the bulk dynamical charges have been re-normalized,
		following the recipe given in Ref.~\onlinecite{Stengel-11}.
	The polarization profile is remarkably flat within the interior of the domains,
	characterized by a constant value
	that amounts to 57.26 $\mu$C/cm$^{2}$ (0.74 $P_{\rm S}$)
	and 45.72 $\mu$C/cm$^{2}$ (0.59 $P_{\rm S}$) for the HH and TT structures, respectively.
	At the domain walls and at the free surfaces we observe a spatial variation of $P(z)$,
	result of localized strong non-homogeneous atomic distortions.
	The extension of these regions with non-uniform polarizations, of approximately six unit cells,
	allow us to quantify the width of the domain walls.
	A more qualitative information can be extracted from a fit of the polarization profile
	to the functional form derived by Gureev {\it et al.}~\cite{Gureev-11},
	
	\begin{equation}
	P(z) = P_{0} \tanh\left( \frac{z}{\delta}\right),
	\label{eq:pzdelta}
	\end{equation}
	
	\noindent where $\delta$ is the typical half-width of the domain wall
	(i.e. to make the transition from $-P_{0}$ to $P_{0}$ we need a space of 
	four times this length $\delta$.
	The results of the fitting of the profiles shown in Fig.~\ref{fig:layer-by-layer-pol}
	to the model summarized in Eq.~(\ref{eq:pzdelta}) can be found in Table~\ref{table:fitpol}.
	\begin{table}
		\caption[ ]{ Results of the fitting of the polarization profile for the HH and TT domain
			walls shown in Fig.~\ref{fig:layer-by-layer-pol} to Eq.~(\ref{eq:pzdelta}).
			$\Delta E$ is the difference in energy of the full slab (162 atoms)
			with respect to the most stable centrosymmetric unpolarized structure.
		}
		\begin{center}
			\begin{tabular}{cccc}
				\hline
				\hline
				&
				$P_{0}$ ($\mu$C/cm$^{2}$) &
				$\delta$ (\AA)            &
				$\Delta E$ (eV)           \\
				\hline
				HH          &
				57.26       &
				5.80        &
				1.374       \\
				TT          &
				45.72       &
				6.87        &
				1.160       \\
				\hline
				\hline
			\end{tabular}
		\end{center}
		\label{table:fitpol}
	\end{table}
	\noindent Clearly, the domain wall widths are larger than in the neutral configuration,
	where they were found to be very narrow, of the order of the lattice constant~\cite{Meyer-02},
	also supporting the conclusions of Ref.~\onlinecite{Gureev-11}.
	
	These 180$^\circ$ domain walls, where the wall is perpendicular to the polarization,
	give rise to polarization-induced bound charges, that lead to the local concentration of a large electrostatic energy that destabilizes this configuration. 
	The stabilization of the domains observed in Fig.~\ref{fig:layer-by-layer-pol} implies
	that the positive (for the HH) and negative (for the TT)
	polarization charges at the domain wall have been screened.
	Since in our calculations we do not consider the presence of defects or 
	dopants~\cite{Wu-06}, the most plausible scenario is the
	metallization of the domain wall and the free surfaces due to the neutralization
	by free carriers, yielding to the formation of thin quasi-two-dimensional metallic
	layers.
	The fingerprint to characterize such scenario would be the presence
	of a substantial amount of free charge populating the band edges 
	(top of the valence band and bottom of the conduction band) both at the 
	domain wall and at the free surfaces of the slab.
	
	\subsection{Layer by layer PDOS}
	\label{sec:pdos}
	
\begin{figure}[h]
		\centering
		\includegraphics [width=\columnwidth]{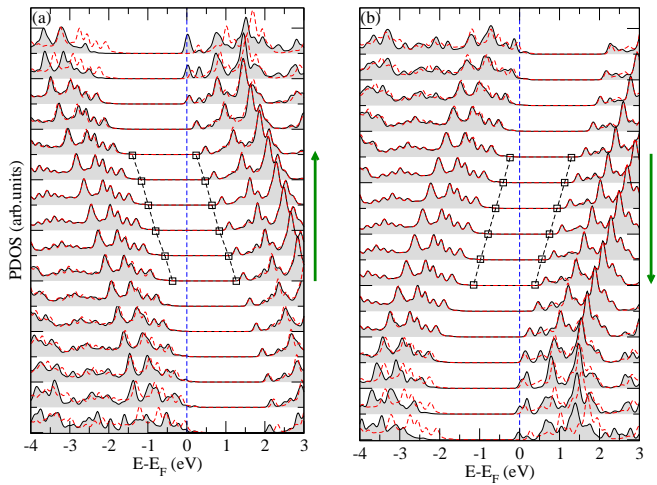}
		\caption {(Color online) Layer by layer PDOS 
			(solid black curves with gray shading) on the TiO$_2$ layers 
			for (a) head-to-head domain, and (b) tail-to-tail 180$^\circ$
			domains. The bottom curve lies at the free surface,
			and the top one lies at the domain wall.
			The bulk PDOS curves (red dashed) are aligned 
			to match the Ti(3$s$) peak. 
			The squares represent the position of the local band edges,
			computed following the recipe of Ref.~\onlinecite{Stengel-11}. 
			The dashed black lines are a linear interpolation of the
			calculated band edges.
			The Fermi level is located at zero energy, 
			as marked by the vertical blue dashed line. 
			Only the PDOS on half of the symmetric supercell, 
			with the polarization direction pointing along the green arrow,
			are shown.}
		\label{fig:pdos}
	\end{figure}

	Fig.~\ref{fig:pdos} shows the layer (spatially) resolved PDOS on all the 
	atomic orbitals of the Ti and O atoms at the TiO$_{2}$
	layers of the different PbTiO$_{3}$ unit cells. 
	The bottom layer plotted in Fig.~\ref{fig:pdos} is adjacent to the free surface
	and the top one lies close to the domain wall. 
	Only the PDOS for one of the domains are shown, since there
	is a mirror symmetry plane at the domain wall.
	(As shown in Ref.~\onlinecite{Stengel-11.2} the equilibrium distribution and spreading 
		of the electron gas is uniquely determined by the boundary values of the electric
		displacement field, that can be indeed set in an asymmetric way. Those asymmetric configurations are beyond the scope of the present work).
	All the PDOS curves were calculated following the recipe
	given in Sec. III.A.1 of Ref.~\onlinecite{Stengel-11} with the Dirac 
	delta functions for the eigenvalues in the PDOS computations 
	replaced by smearing normalized Gaussians with a finite 
	width that is twice as large, as suggested in Appendix B
	of Ref.~\onlinecite{Stengel-11}.
	On top of the heterostructure PDOS we superimpose the bulk
	PDOS (dashed red lines), calculated with an equivalent $k$-point sampling.
	The bulk reference calculation was computed from a periodic bulk calculation
	with a five atom per unit cell PbTiO$_{3}$ structure,
	with the atomic distortions and out-of-plane strain extracted from a region of the 
	HH or TT domains where the relaxed atomic
	structure (Fig.~\ref{fig:layer-by-layer-pol}) has converged into a regular pattern. 
	Then we extract the PDOS for the same TiO$_{2}$ atomic layer.
	Finally,we note that the superposition of the bulk PDOS and supercell PDOS at each layer was done by matching the sharp peaks of the Ti(3$s$) semicore band. 
	
	First of all, it is remarkable to check how for the perovskite material
	studied in this work, PbTiO$_{3}$, and even upon the electronic reconstruction and metallization, 
	a well-defined energy gap persists in all layers between the bands that are mostly O-$p$ in character
	(top of the valence band at the bulk level), and the bands that are mostly 
	Ti-$t_{\rm 2g}$ in character (bottom of the conduction band at the bulk level).

	The local metallization of the free-surface layers and the domain wall
	is evident from the location of the valence-band top and the conduction-band bottom 
	of these layers with respect to the Fermi level of 
	the whole structure, taken as zero in Fig.~\ref{fig:pdos}.
	For the HH domain configuration [Fig.~\ref{fig:pdos}(a)], 
	the non-vanishing PDOS at the Fermi level for the three topmost TiO$_{2}$ layers 
	points to the electronic population of the conduction-band bottom of the PbTiO$_{3}$ unit cells
	close to the domain wall.
	This free electronic charge has been transferred from the free surface, 
	as can be clearly seen from the generation of holes at the bottom-most layers
	(see the crossing of the Fermi level with the PDOS at the valence-band top for the
	three bottom-most TiO$_{2}$ layers).
	For the TT domain configuration [Fig.~\ref{fig:pdos}(b)],
	the same behaviour is found, but with the role of the electrons and holes interchanged
	(i.e. generation of holes at the top of the valence band in the layers 
	adjacent to the domain wall, and the population of the bottom of
	the conduction band with electrons close to the free surfaces).
	
	The PDOS of the conduction and valence bands converges
	fairly quickly to the bulk curve when moving away from
	the interface, and the PDOS vanish at the Fermi level, which implies
	that the system is locally insulating. 
	Furthermore, the PDOS in each layer appears rigidly shifted with respect
		to the neighboring two layers, consistent with the presence
		of an internal remnant depolarization field that keep the free charges electrostatically confined
		to the metallic region within the domains~\cite{Stengel-11} (see black dashed lines in Fig.~\ref{fig:pdos}).
	Since the bulk calculations were performed at zero external electric field,
	while in the slab calculations there is a residual depolarizing field,
	the coincidence of the two curves far enough from the free surfaces and the
	domain wall means that the main effect on the layer by layer PDOS comes from the
	lattice distortions, and not from macroscopic depolarizing fields.
	This residual depolarizing field will play a major role in the determination
	of the atomic orbitals involved in the electronic reconstruction, as will be discussed below.

	\subsection{Coupling between density of free carriers and the polarization profiles}
	\label{sec:charge carriers}
	\begin{figure}[h]
		\centering
		\includegraphics [width=\columnwidth]{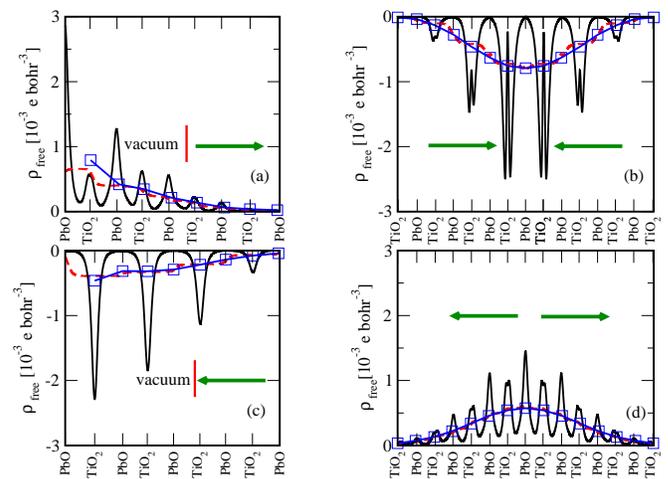}
		\caption {(Color online) Calculated free charge for head-to-head and tail-to-tail domain walls. 
			The solid black curve represents the planar averaged $\overline{\rho}_{\rm free}$, as defined
			in Eq.~(\ref{eq:planaraverage}), and the red dashed line corresponds to the nanosmoothed
			average computed as in Ref.~\onlinecite{Junquera-07}.
			The blue line represents the profile of the magnitude of the bound charge, computed from the finite difference
			derivative of the layer-by-layer polarization shown in Fig.~\ref{fig:layer-by-layer-pol}. The sign of this bound charge opposes  to that of the free charge, both at the surface and at the domain wall.
			(a) Hole charge density at the free surface, and (b) electron charge density at the domain wall
			for the head-to-head configuration.
			(c) Electron charge density at the free surface, and (d) hole charge density at the domain wall 
			for the tail-to-tail configuration.
			The green arrows represent the polarization configuration in each case.
			Negative densities correspond to electron gases and positive densities for hole gases.}
		\label{fig:chargy}
	\end{figure}
	
	To estimate the amount of charge that has been transferred from the free-surfaces
	to the domain wall in order to screen the depolarizing fields,
	we plot in Fig.~\ref{fig:chargy} the planar average of the free charge,
	$\overline{\rho}_{\rm free} (z)$, defined as 
	
	\begin{equation}
	\overline{\rho}_{\rm free} (z) = \frac{1}{S} \int_{S} \rho_{\rm free} ({\bf r}) dx dy,
	\label{eq:planaraverage}
	\end{equation}
	
	\noindent where $S$ is the area of the interface unit cell,
	and the corresponding nanosmoothed version~\cite{Junquera-07}, 
	$\overline{\overline{\rho}}_{\rm free}(z)$.
	For the electronic charge density populating the bottom of the
	conduction band $\rho_{\rm free} ({\bf r})$ has been computed as
	
	\begin{equation}
	\rho_{\rm free}^{\rm electrons} ({\bf r}) = \sum_{E_{n \bf{k}}>E_{0}} w_{\bf{k}} f_{n\bf{k}} \vert \psi_{n \bf{k}} ({\bf{r}}) \vert^{2},
	\end{equation}
	
	\noindent while for the hole charge density populating the top of the valence band, it amounts to
	
	\begin{equation}
	\rho_{\rm free}^{\rm holes} ({\bf r}) = \sum_{E_{n \bf{k}}<E_{0}} w_{\bf{k}} (1 - f_{n\bf{k}}) \vert \psi_{n \bf{k}} ({\bf{r}})\vert^{2},
	\end{equation}
	
	\noindent where $w_{\bf{k}}$ is the weight of a special $\bf{k}$ point in the discrete 
	Monkhorst-Pack mesh, $f_{n \bf{k}}$ are the occupation numbers of the eigenstate $\psi_{n \bf{k}}$ and
	the sum is restricted to the states with eigenvalue $E_{n \bf{k}}$ higher than $E_{0}$, a value of the energy
	corresponding to the center of the gap between valence and conduction band.
	The results are shown in Fig.~\ref{fig:chargy}.
	
	Again, we clearly see that the system is not locally charge
	neutral at the domain wall and at the free-surfaces.
	If the profiles of the nanosmoothed average free charge densities 
	are integrated along $z$, we can extract the free-carrier
	per unit area, labeled as $\sigma_{\rm e}$ (for the electrons)
	and $\sigma_{\rm h}$ (for the holes).
	For the HH domains, there is an additional 
	electron density at the domain wall populating the
	bottom of the conduction bands that amounts to $\sigma^{\rm HH}_{\rm e}$ = 1.033 electrons
	[Fig.~\ref{fig:chargy}(a)].
	Those electrons are transferred from the free-surfaces,
	where the integration of the free hole charge in 
	Fig.~\ref{fig:chargy}(b) yields a value of $\sigma^{\rm HH}_{\rm h}$ = 0.517 holes per surface unit cell.
	As required by the global charge neutrality $\sigma^{\rm HH}_{\rm e} = 2 \sigma^{\rm HH}_{\rm h}$
	(we have two free surfaces populated with holes and just one HH domain wall
	charged with electrons).
	
	For the TT domains, the extra holes at the
	domain wall [Fig.~\ref{fig:chargy}(c)] equals $\sigma^{\rm TT}_{\rm h}$ = 0.827, that 
	again just double the number of extra electrons populating
	the bottom of the conduction bands at the free-surfaces, $\sigma^{\rm TT}_{\rm e}$ = 0.413
	[Fig.~\ref{fig:chargy}(d)].
	
\begin{figure}[h]
		\centering
		\includegraphics [width=\columnwidth]{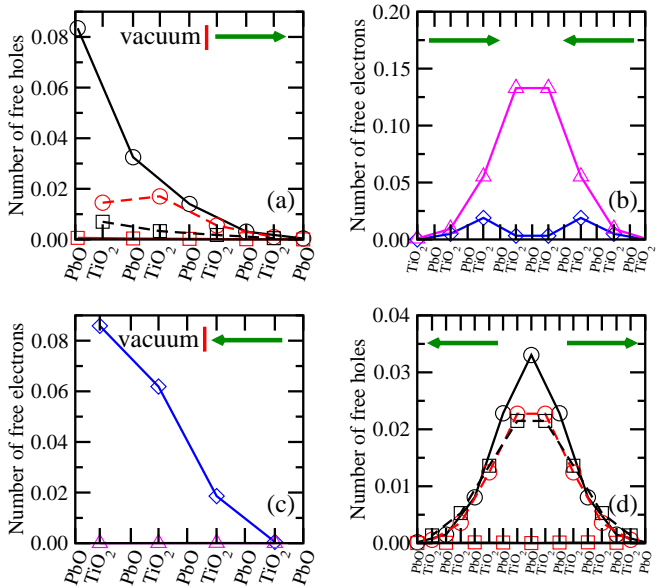}
		\caption {(Color online) Free carrier densities in PbTiO$_{3}$ for
			(a) the free surface and (b) the domain wall in the HH structure,
			(c) the free surface and (d) the domain wall for the TT structure.
			Black circles (respectively, red squares) represent the amount of free holes
			in O orbitals responsible of the $\pi$-bonding (respectively, $\sigma$-bonding).
			The dashed lines indicate that the O are located in a TiO$_{2}$ plane and 
			the solid lines are used for O in the PbO layers.
			Blue diamonds (respectively, magenta triangles up) represent the 
			amount of free electrons in the $d_{xy}$ (respectively $d_{xz}+d_{yz}$) 
			orbitals of Ti. The green arrows represent the polarization configuration in each case.
		}
		\label{fig:decom-orb}
	\end{figure}

	A further insight on the origin of these electron and hole free charges can 
	be drawn after its decomposition into different orbital components,
	as it is done in Fig.~\ref{fig:decom-orb}. 
	The current wisdom is that the $d_{xy}$ orbitals are the first to be populated in
	the related LaAlO$_{3}$/SrTiO$_{3}$~\cite{Popovic-08} or PbTiO$_{3}$/SrTiO$_{3}$~\cite{Yin-15} 
	interfaces, where two-dimensional electron gases are formed to 
	screen the polar catastrophe. 
	This situation is similar to that found in bulk polarized 
	PbTiO$_3$ and also observed in the free surface of our system [Fig.~\ref{fig:decom-orb}(c)].
	However, when discussing the metallic states formed around the domain walls we can observe, on the one hand, that the HH electron gas (Fig.~\ref{fig:decom-orb}b) is, for the most part, formed by $d_{xz}$ and $d_{yz}$ orbitals, while $d_{xy}$ ones host very little charge. This can also be observed in the doubled-peaks of Fig.~\ref{fig:chargy}b.
	On the other hand, the TT hole gas has a primary contribution  from the $\pi$-orbitals in the PbO plane 
	(i.e. $p_{x}$ and $p_{y}$ orbitals in the oxygen at the PbO layer)
	and a smaller, but also significant, contribution from both $\sigma$ (i.e. the $p_{x}$, respectively
	$p_{y}$, orbital of the equatorial O located along $x$, respectively $y$, with respect the Ti) 
	and $\pi$-bonded orbitals (i.e. the $p_{y}$ and $p_{z}$, respectively
	$p_{x}$ and $p_{z}$, orbitals of the equatorial O located along $x$, respectively $y$, with respect the Ti)
	in TiO$_2$ planes. 
	Both facts are consistent with the effects of localization due to the remnant electric potential energy
	within the domains, whose shape resembles a $\vee$ for the HH and $\wedge$ for the TT, with the peak located right
	at the domain wall in both cases.
	Those are confinement potentials for the respective carriers.
	For the bands with majority weights on orbitals directed parallel to the domain wall, there is no effect
	on the dispersion; only the center of the subbands might be displaced following exactly the shape of the potential.
	This is the case of the $d_{xy}$ bands of Ti, and on the O-$\sigma$ orbitals.
	However, the bands with majority weights on orbitals directed perpendicular to the domain wall, both the 
	dispersion and the local center of mass of the bands are strongly modified due to the confinement potential:
	the range of the possible interactions is strongly reduced.
	As a consequence the hybridization is decreased, and the corresponding bonding orbitals 
	(O-$\pi$ orbitals in PbO layers) increase the
	energy and the antibonding orbitals ($d_{xz}$,$d_{yz}$) suffer a concomitant reduction.
	Those are the first orbitals to be filled with electrons [Fig.~\ref{fig:decom-orb}(b)] or emptied 
	to create holes [Fig.~\ref{fig:decom-orb}(d)] around the domain wall.
	
	In order to elucidate how this electronic reconstruction
	is a result of the screening of the polarization-induced
	charges at the 180$^\circ$ domain walls, we also plot
	in Fig.~\ref{fig:chargy} a numerical differentiation of the
	polarization profile shown in Fig.~\ref{fig:layer-by-layer-pol}.
	There is an essentially perfect matching between the nanosmoothed free charge densities
	and the bound charge profile, $\rho_{\rm b}$ (blue lines in Fig.~\ref{fig:chargy}),
	defined as $\rho_{\rm b} (z) = - dP/dz$. 
	Therefore, the polarization discontinuity at the domain wall is the driving force
	for the electronic reconstruction, and this transfer of charge between the domain wall and
	the free surface provides an effective screening mechanism that stabilizes
	the monodomain configurations in both domains.
	
	\section{Conclusions}
	\label{sec:conclusions}
	
	Our density-functional simulations have shown how 180$^\circ$ HH and TT charged domains can be
	stabilized in the out-of-plane monodomain configuration in an ideal PbTiO$_{3}$ slab
	(no dopants nor vacancies), thanks to the electrostatic screening provided by free carriers.
	Above a critical size of the domains, in our case
	$n$ = 16 unit cells, an electronic reconstruction
	takes place, where charge is transferred from the
	free-surface of the slab to the domain wall,
	and generates quasi-two-dimensional electron 
	and hole gases.
	The width of the domain wall of around 6 (for the HH) and 7 (for the TT) unit cells, is much larger than the typical width of neutral domains (of the order of one lattice constant).
	A clear electrostatic coupling between the density of free-charge and the polarization profile, whose negative gradient equals the bound charge, is found.
	
	Our results support previous phenomenological models based on Landau and semiconductor theory, regarding the existence of quasi-two-dimensional charge densities at the 180$^\circ$ HH-TT
	domain walls in isolated ferroelectric~\cite{Gureev-11}, and postulated them to be the driving force for the 
	metallic like conductivity found at the walls between insulating polar materials~\cite{Sluka-13}.
	
	The driving force for the accumulation of free charges at the domain wall 
		is the mismatch of the formal polarization between the constituent materials~\cite{Aguado-Puente-15,Yin_15}.
		(This formal polarization corresponds to the raw result of a Berry phase calculation~\cite{Stengel-09}.)
		Therefore, in the quest of searching for potential domain walls where this scenario can occur, the larger the polarization mismatch, the better.
		In a ferroelectric material, the formal polarization can be tuned with 
		electric fields, strain, or temperature,
		offering more degrees of freedom than the celebrated two-dimensional electron gas at the
		SrTiO$_{3}$/LaAlO$_{3}$ interface~\cite{Ohtomo-04}, where the formal polarization
		in LaAlO$_{3}$ is due to the 
		fact that the individual LaO and AlO$_{2}$ layers have formal charges of $\pm e$~\cite{Stengel-09}.
		Also, it is important to note that the smaller the band gap of the material, the shorter the critical thickness 
		required by the electrostatic potential to close the gap, and the sooner 
		the charged domain walls would appear.
	
	These first-principles simulations might serve as benchmarks to perform second-principles computations
	that include all the relevant electronic and lattice degrees of freedom~\cite{Garcia-16} in very
	large systems under operating conditions of field and temperature.
	Once the atomistic and electronic models for the 
	second-principles calculations will be validated,
	then more challenging calculations including tilting domains like the ones that appear during domain switching, or the conductivity at domain walls can be tackled.

	\section*{ACKNOWLEDGMENT} 
	\label{sec:Acknowledgement} 
	We acknowledge Pablo Aguado-Puente for the  useful discussions we had during the course of this work. J.S. thanks the University of Cantabria for the scholarship funded by the Vice-rectorate for Internationalisation and the Theoretical Condensed Matter Group. We gratefully acknowledge the African school for Electronic Structure Methods and Applications (ASESMA), especially Richard Martin, for assisting in bringing us together to form our collaboration. J.J. and P.G.-F. acknowledge financial support from the Spanish Ministry of Economy and Competitiveness through the MINECO Grant No. FIS2015-64886-394-C5-2-P, and the Spanish Ministry of Science, Innovation and Universities through the grant No. PGC2018-096955-B-C41. P.G.-F. acknowledges support from Ram\'on y Cajal Grant No. RyC-2013-12515. G.S.M acknowledges support from the Kenya Education Network through CMMS mini-grant 2019/2020. The authors also gratefully acknowledge the computer resources, technical expertise, and assistance provided by the Centre for High Performance Computing (CHPC), Cape Town, South Africa. 
	
\bibliography{bibliography}	

\begin{thebibliography}{40}
\bibitem [{bas()}]{basis-request}%
  \BibitemOpen
  \href@noop {} {}\bibinfo {note} {Values for the parameters can be obtained
  upon request.}\BibitemShut {Stop}%
\end{thebibliography}
\end{document}